\documentclass[twocolumn, switch]{article}

\usepackage{PRIMEarxiv}
\usepackage[utf8]{inputenc}
\usepackage[T1]{fontenc}
\usepackage{amsmath, amsthm, amssymb, amsfonts}
\usepackage{nicefrac}
\usepackage{hyperref}
\usepackage{url}
\usepackage{booktabs}
\usepackage{microtype}
\usepackage{fancyhdr}
\usepackage{graphicx}
\usepackage{xcolor}
\usepackage{float}
\usepackage{enumitem}
\usepackage{multirow}

\title{Latent Neural Cellular Automata for Resource-Efficient Image Restoration}

\newcommand{\email}[1]{\href{mailto:#1}{\texttt{#1}}}
\author{
    Andrea Menta\\
    DEIB, Politecnico di Milano\\
    Via Ponzio 34/5, 20133 Milan, Italy\\
    \email{andrea.menta@polimi.it} \\
    \And
    Alberto Archetti\\
    DEIB, Politecnico di Milano\\
    Via Ponzio 34/5, 20133 Milan, Italy\\
    \email{alberto.archetti@polimi.it} \\
    \And
    Matteo Matteucci\\
    DEIB, Politecnico di Milano\\
    Via Ponzio 34/5, 20133 Milan, Italy\\
    \email{matteo.matteucci@polimi.it} \\
}

\begin{document}

\twocolumn[
  \begin{@twocolumnfalse}

\maketitle

\begin{abstract}
Neural cellular automata represent an evolution of the traditional cellular automata model, enhanced by the integration of a deep learning-based transition function. This shift from a manual to a data-driven approach significantly increases the adaptability of these models, enabling their application in diverse domains, including content generation and artificial life. However, their widespread application has been hampered by significant computational requirements. In this work, we introduce the Latent Neural Cellular Automata (LNCA) model, a novel architecture designed to address the resource limitations of neural cellular automata. Our approach shifts the computation from the conventional input space to a specially designed latent space, relying on a pre-trained autoencoder. We apply our model in the context of image restoration, which aims to reconstruct high-quality images from their degraded versions. This modification not only reduces the model's resource consumption but also maintains a flexible framework suitable for various applications. Our model achieves a significant reduction in computational requirements while maintaining high reconstruction fidelity. This increase in efficiency allows for inputs up to 16 times larger than current state-of-the-art neural cellular automata models, using the same resources.
\end{abstract}
\vspace{0.75cm}

  \end{@twocolumnfalse}
]



\section{Introduction}
\label{se:introduction}

Cellular Automata (CA)~\cite{neumann1951} are a computational model for simulating the dynamics of complex systems from simple local interactions. CAs operate on a regular lattice in which each cell is in one of a finite number of states. At each discrete time step, cells are updated using simple local rules based on the state of neighboring cells. CAs are capable of generating complex patterns from simple update rules. The most famous example is John Conway's Game of Life, which is Turing complete~\cite{rendell2002}. In general, the potential to generate intricate global patterns from simple local rules has always attracted interest in CA research.

The advent of Deep Learning (DL)~\cite{goodfellow2016} has revolutionized various domains, such as computer vision and natural language processing, establishing itself as a State-of-the-Art (SotA) methodology. Its inherent flexibility and generalization power have allowed several extensions of existing models, including CAs. Neural Cellular Automata (NCA)~\cite{mordvintsev2020} exemplify this progression, by integrating DL into CAs.
NCAs differ from CAs by using neural networks as their transition functions. This aspect allows the transition from manually designed rules to data-driven learning processes that target specific behaviors. Although many promising applications have been proposed, ranging from content generation~\cite{sudhakaran2021} to control~\cite{variengien2021}, NCAs face significant limitations due to their resource requirements. The need for numerous concurrent updates introduces significant latency and memory overhead~\cite{tesfaldet2022}, similar to Recurrent Neural Networks (RNNs)~\cite{rumelhart1986_1}, hindering their practical application. 

To address these challenges, we introduce the Latent Neural Cellular Automata (LNCA) model. The core of LNCA is to move computations from the conventional input space to a customized, compressed latent space. The idea is to distill the essential information for the NCA to solve the problem into a lower-dimensional manifold, leaving the accessory information to a secondary component.  For this purpose, an Autoencoder (AE) is used to compress the input and learn an embedding function.
Our research focuses primarily on Image Restoration (IR), tackling denoising and deblurring tasks. We evaluate LNCA against SotA solutions such as Restormer~\cite{zamir2022} and NAFNet~\cite{chen2022}, which are based on standard deep learning techniques, as well as the best NCA model, called ViTCA~\cite{tesfaldet2022}. Our evaluation includes 10 datasets, both real and synthetic, measuring reconstruction performance and computational efficiency.

To the best of our knowledge, LNCA is the first model applying NCAs in the latent space of a DL autoencoder backbone. Thus, the contributions of this work are twofold:
\begin{itemize}[leftmargin=0.5cm]
\item We propose a novel architecture called LNCA and the corresponding learning process to mitigate the inherent limitations of NCA models by exploiting the expressiveness of DL autoencoders.
\item We apply the LNCA architecture to IR tasks, achieving significant reductions in latency and GPU memory usage while maintaining competitive reconstruction performance compared to SotA models.
\end{itemize}
To facilitate further research, we have made the source code of the experiments publicly available.

\section{Related Work}
\label{se:related_work}

This section describes the main works related to (neural) cellular automata and image restoration.

\subsection{Cellular Automata}
\label{subse:cellular_automata}
Cellular Automata (CA) are defined as a tuple $\mathcal{M} = (\mathcal{L}, \mathcal{S}, \eta, \phi)$, where $\mathcal{L}$ is a $d$-dimensional lattice, $\mathcal{S}$ is the set of possible states, $\eta:\mathcal{L}\rightarrow 2^\mathcal{L}$ is the neighborhood function and $\phi:\mathcal{S}^n\rightarrow\mathcal{S}$ is the transition function.
Let $s^t_i\in\mathcal{S}$ indicate the state of cell $i$ at time step $t$ and $\Omega^t_i=\left\{s^t_j : j\in\eta(i)\right\}$ the set of neighbors of cell $i$ according to $\eta$.
Then, the state update for cell $i$ at each iteration $t$ is computed as
\begin{equation}
\label{eq:update_rule}
    s_i^{t+1}=\phi\left(\left\{s_i^t \cup \Omega_i^t\right\}\right)\text{.}
\end{equation}

Neural Cellular Automata (NCAs) extend the CA framework by replacing the transition function $\phi$ with a learnable neural network $\phi_\theta$ with parameters $\theta$, preserving the essential feature of locality. However, training the neural network $\phi_\theta$ requires some modifications to standard CAs. First, the state representations are continuous. Second, optimization is performed after $n$ processing iterations, and a boolean mask is used during each iteration to allow asynchronous updates, which improves training dynamics, as in ~\cite{mordvintsev2020}. This training methodology typically involves a two-step process with a replay buffer to perform pool sampling. During each odd step, the neural network updates a newly initialized tensor $x$, and the resulting tensor $x'$ is stored in the replay buffer. For even steps, instead, the network retrieves $x'$ from the replay buffer, updates it, and then stores the new tensor $x''$ back in the buffer. This strategy ensures that the target output acts as an attractor in the training process, effectively preventing the neural network from deviating from the desired behavior.


\subsection{Image Restoration}
\label{subse:stat_of_the_art_of_image_restoration}

Image Restoration (IR) is a central task in digital image processing that aims to reconstruct corrupted images into their original form. This process involves reducing distortions such as noise and blur. Over time, a plethora of methods have been developed to tackle IR, ranging from basic filtering techniques~\cite{banham1997} to advanced deep learning models, each improving the quality of the restoration. In our analysis, we focus on three state-of-the-art deep learning-based architectures, called Restormer, NAFNet, and ViTCA.

Restormer~\cite{zamir2022}, leverages a residual learning approach based on the classic UNet~\cite{ronneberger2015unet} architecture, powered by Transformer blocks~\cite{vaswani2017attention}. Each block is composed of a Multi-Dconv Head Transposed Attention (MDTA) and a Gated-Dconv Feed-forward Network (GDFN). MDTA offers a memory-efficient alternative to the self-attention block by utilizing channel dimension, while GDFN incorporates a gating mechanism to the classic transformer head.

NAFNet~\cite{chen2022} is a non-linear activation-free network, based on a UNet structure. The basic block of this architecture includes two components, the Simple Gate (SG) and the Simplified Channel Attention (SCA). SG introduces non-linearities in the block, while SCA performs an attention operation leveraging a channel-wise product. Both SG and SCA derive from the gated linear unit~\cite{dauphin2017language}.

Finally, ViTCA~\cite{tesfaldet2022} is the first application of the attention mechanism to the NCA family. This CA-based architecture implements a transition function composed of four stages: embedding, Multi-Head Self-Attention (MHSA), Multi-Layer Perceptron (MLP), and head. 
The embedding takes the seeded input (optionally including the positional encoding) and feeds it through a linear projection.
The MHSA, the major novelty of ViTCA, estimates the relative importance of each token.
It implements a localized self-attention method to respect NCA locality constraints, limiting the attention span to each cell’s neighborhood. 
This way, even though information propagates locally, it achieves a global behavior.
The MLP, composed of two fully connected layers with GELU activation, completes the classic Vision Transformer architecture~\cite{dosovitskiy2021}, while the head reshapes the output obtaining the final update vector. 
Lastly, this model is trained using pool sampling and curriculum learning, a typical approach for NCA architectures.

\section{Latent Neural Cellular Automata}
\label{se:latent_neural_cellular_automata}

This paper introduces a novel approach to address the resource constraints associated with NCAs, called Latent Neural Cellular Automaton (LNCA). The core concept of our methodology is to move the NCA computation~-- local information passing and state transition~-- from the input space to a specially designed latent space. The main goal motivating this shift is to limit the dimensionality of the space in which the NCA operates, extracting only the task-specific information required and reducing redundant operations. In the context of IR, our approach is particularly advantageous. We aim to build an embedding space featuring a distinct separation between the image semantics and the information about its corruption. This way, the autoencoder focuses on constructing an embedding space structured to decouple image and corruption information, while the NCA performs the restoration task. This task division significantly enhances the overall efficiency of the system. In the following, we present the LNCA architecture and its training pipeline within the IR task.

\subsection{Architecture}
\label{subse:architecture}

\begin{figure*}
    \centering
    \includegraphics[width=\textwidth]{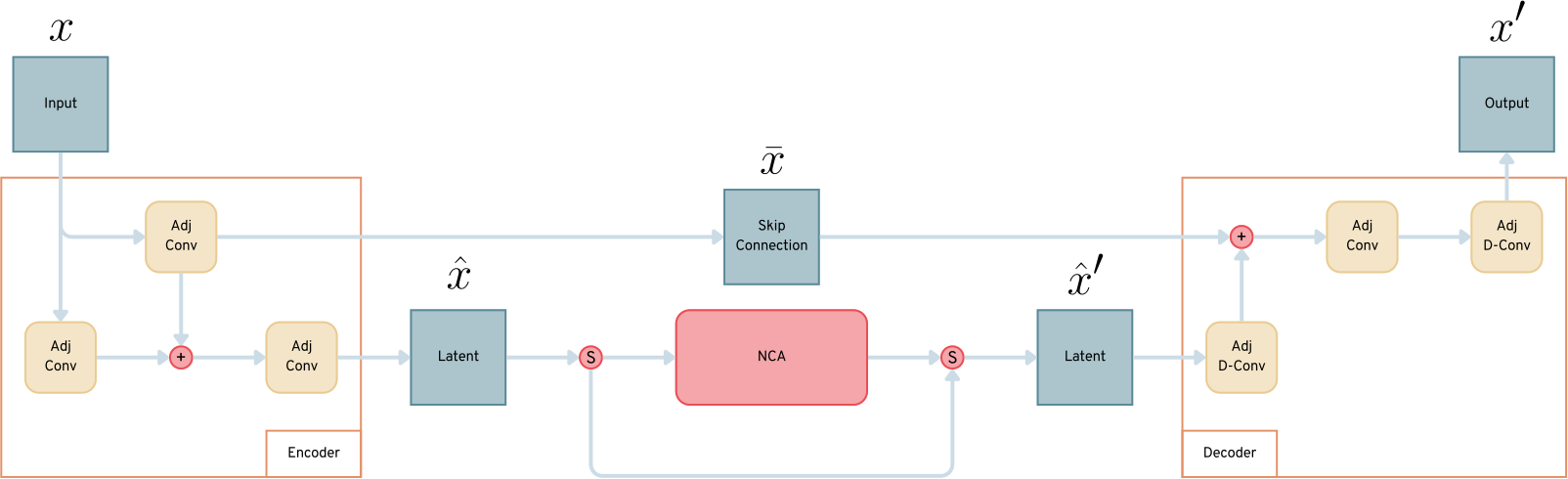}
    \caption[LNCA structure]{LNCA structure. $+$ is an element-wise addition. $S$ is a channel switch. On the left, the Encoder, which is composed of adjusted convolution blocks, outputs the skip-connection $\bar{x}$ and the latent tensor $\hat{x}$.
    In the middle, the NCA block processes the latent tensor to remove the corruption, obtaining $\hat{x}^\prime$.
    On the right, the Decoder uses both the skip-connection and the latent tensor to reconstruct the final output $x^\prime$.
    The switch path around the NCA is used to bypass the NCA computation during the AE training, as described in Section~\ref{subsubse:autoencoder_training}.}
    \label{fig:lnca_structure}
\end{figure*}

The LNCA architecture is composed of two modules, the Autoencoder (AE) and the Neural Cellular Automaton (NCA). The AE's encoder module projects input images from the input space to the latent space, while the AE's decoder performs the inverse transformation. The NCA, instead, performs the actual image restoration operating in the latent space. Figure~\ref{fig:lnca_structure} shows the LNCA architecture. Specifically, the input image $x\in\mathbb{R}^{H\times W\times C}$ is translated by the encoder $\mathcal{E}$ of the AE into its latent representation $\hat{x}\in\mathbb{R}^{\hat{H}\times \hat{W}\times \hat{C}}$. Then, the NCA $\mathcal{N}$ performs $n$ transition steps starting from the initial state $\hat{x}$ and obtaining a new latent representation $\hat{x}^\prime$ with the same dimensionality. Finally, the decoder block $\mathcal{D}$ of the AE transforms the latent denoised image $\hat{x}^\prime$ to the original image space, obtaining the desired output $x^\prime\in\mathbb{R}^{H\times W\times C}$.

\subsubsection{Autoencoder}
\label{subsubse:autoencoder}

The AE's core objective is to create a lower dimensional manifold of the input space containing task-related information for the NCA while preserving reconstruction fidelity. For this reason, our architecture opts for a classical AE model instead of more sophisticated variants, such as variational autoencoders~\cite{kingma2014vae}, which may introduce additional reconstruction uncertainty. As shown in Figure~\ref{fig:lnca_structure}, we adopted a standard AE architecture with some minor tweaks to obtain better convergence during training. First, we included a variant of the convolutional block, called adjusted convolution. This block comprises an initial convolution with $3\times3$ kernel and no activation, followed by batch normalization, and swish activation function. The only exception is the output layer which uses a sigmoid activation. The same is applied to the transposed version of convolution used in the decoder. In general, this block grants better convergence properties and stability for the AE.

To help the latent space construction, we made two major additions to the AE architecture, namely a skip-connection and an encoder side-channel. As skip-connections aid in reconstruction quality, they funnel image semantics unrelated to the corruption directly to the decoder. At the same time, all the restoration information flows into the AE's bottleneck for NCA processing. This behavior is enforced through a composite loss function, detailed in Section~\ref{subsubse:autoencoder_training}.

\subsubsection{Neural Cellular Automata}
\label{subsubse:neural_cellular_automata}

The NCA is the component devoted to the actual restoration task. We opted for integrating a ViTCA-based NCA as the NCA core of our architecture, as it stands as the SotA model within the NCA family. Additionally, this enables a direct comparison with the original, non-latent ViTCA and our novel model in terms of reconstruction performance and computational complexity.

To further reduce computational requirements and evaluate the impact of the attention mechanism for NCA architectures, we developed an alternative attention-less NCA model.
This model, called NAFCA, is inspired by NAFNet, the architecture that replaced self-attention with a lighter gating mechanism, adjusted to satisfy the CA locality constraint.
This is achieved by substituting the transformer-based transition function with a simpler architecture.
Specifically, this model comprises three residual blocks: the perception block, the update block, and the head. 
The perception block performs the information extraction and it is composed of an initial layer normalization, followed by an activation-free $1\times1$ convolution and a Simple Gate~\cite{chen2022}. 
The classic attention mechanism is then substituted by the following operation:
\begin{equation}
\label{eq:nafca_attention}
    x^\prime = x\circ\mathcal{C}(x)\text{,}
\end{equation}
where $\mathcal{C}$ is a single-filter $3\times3$ convolution respecting the Moore neighborhood and $\circ$ is the Hadamard product. 
Then, another $1\times1$ convolution and a dropout layer complete the perception block. 
The update block takes care of the transition process. 
Its structure is similar to the perception block, except for the lack of attention mechanism and the number of filters in the convolutional layers. 
Finally, the head handles the state update. 
It is composed of a layer normalization and a $1\times 1$ convolution.

\subsection{Training}
\label{subse:training}

In the context of NCA, the training procedure must be carefully designed to allow model convergence. 
We chose a training strategy consisting of two phases to allow for more robust convergence. 
The first phase is the AE training, where the AE learns to build a latent space suitable for the NCA while minimizing the reconstruction error. 
The second phase is NCA training, where the NCA learns to reconstruct the image within the latent space. 
In each phase, the non-trained part of the architecture (i.e., first the NCA, then the AE) is kept frozen.
This two-phase approach allows a clear separation of tasks between the AE and the NCA, driving the architecture blocks to the desired behavior, and avoiding contamination.

\subsubsection{Autoencoder Training}
\label{subsubse:autoencoder_training}

The first training phase involves the AE block only. Thus, the NCA processing is bypassed with a direct connection between the encoder and the decoder. This phase leverages batches $x$ composed of three types of samples: 
\begin{itemize}[leftmargin=0.5cm]
    \item Anchor samples ($x_A$), the ground truth restored images.
    \item Positive samples ($x_P$), corrupted versions of the ground truth.
    \item Negative samples ($x_N$), derangement (permutation where no element appears in its original position) of the ground truth.
\end{itemize}
Given a sample batch, the AE training procedure is composed of two separate gradient descent steps with dedicated losses to ensure more stable convergence. 

The first step includes the sum of three losses: the reconstruction loss, the distance loss, and the task loss. The goal of the reconstruction loss is to minimize the AE's reconstruction error and it is evaluated as the mean squared error between the AE's input and output tensor:
\begin{equation}
\mathcal{L}_{\text{REC}_{\text{AE}}}(x, x^\prime) = \text{MSE}(x, x^\prime)\text{.}
\label{eq:ae_training_reconstruction_loss_1}
\end{equation}
The distance loss takes inspiration from the Triplet loss~\cite{schroff2015}  and its goal is to enforce the similarity in the latent space between images with the same subjects and the dissimilarity between images with different subjects. Given a tolerance margin $\alpha$, this loss is calculated as
\begin{equation}
\mathcal{L}_{\text{DIST}}(\hat{x}) = \max \{0, \text{MSE}(\hat{x}_A, \hat{x}_P) - \text{MSE}(\hat{x}_A, \hat{x}_N) + \alpha \}\text{.}
\label{eq:ae_training_distance_loss_1}
\end{equation}
Finally, the task loss is used to obtain a faithful reconstruction of the input corruption. To enforce this property, this loss takes into account only the corrupted pixels. Specifically, it applies a boolean mask before computing the mean squared error of the corrupted samples:
\begin{equation}
\begin{split}
\mathcal{L}_{\text{TASK}}(x, x^\prime) & = \text{MSE}(x_P\circ m_{A,P}, x_P^\prime\circ m_{A,P}) \\
m_{A,P} & = \mathbf{1}(x_A - x_P \neq 0)\text{.}
\end{split}
\label{eq:ae_training_task_loss_1}
\end{equation}
Each of these losses is then weighted and summed to obtain the loss of the first gradient descent step.
The weights are determined through validation over a finite set of values using grid-search.
This tuning procedure ensures a balance between reconstruction quality and latent space structure and helps manage differences in loss magnitudes.

The second gradient descent step, instead, involves the equivalence loss, inspired by the Swapping AE~\cite{park2020} design. The goal of this loss is to split the information related to the image semantics and the image corruption between the two encoder outputs.
In particular, the skip-connection $\bar{x}$ must help in the reconstruction of the image semantics, while the corruption profile is encoded only in the latent tensor $\hat{x}$.
This is obtained by swapping $\hat{x}_P$ and $\hat{x}_A$, the latent tensors corresponding to the corrupted images and the ground truth images, respectively.
In fact, by substituting $\hat{x}_P$ with its clean counterpart $\hat{x}_A$, while keeping the corrupted skip-connection $\bar{x}_P$, we should obtain a clean output $x_A$ and vice versa.
This approach has the advantage of introducing a specific behavior in the AE, which is essential for the subsequent training procedure of the NCA.
We enforce this property by minimizing the relative mean squared error: 
\begin{equation}
\begin{split}
\mathcal{L}_{\text{EQ}}(x, \bar{x}, \hat{x}) = & \, \text{MSE}(x_A \circ m_{A,P}, x^\star_A \circ m_{A,P}) + \\
& \,\text{MSE}(x_P \circ m_{A,P}, x^\star_P \circ m_{A,P}) \\
x_A^\star = & \, \mathcal{D}(\hat{x}_A + \epsilon, \bar{x}_P) \\
x_P^\star = & \, \mathcal{D}(\hat{x}_P, \bar{x}_A)\text{,}
\end{split}
\label{eq:ae_training_equivalent_loss_1}
\end{equation}
where $\epsilon\sim N(0,v)$ is a smoothing factor that allows for a small tolerance in the reconstruction.

\subsubsection{Neural Cellular Automata Training}
\label{subsubse:neural_cellular_automata_training}
In this phase, we focus on the restoration process.
Each input batch $x$ is composed only of corrupted images and it is associated with a clean ground truth $y$. 
Furthermore, the generated latent tensor $\hat{x}$ is processed by the NCA module, obtaining $\hat{x}^\prime$. 

Similar to the AE training, this procedure leverages multiple losses to learn the IR task: the reconstruction loss, the latent loss, and the overflow loss.
Each loss has an associated weight, which is tuned similarly to the AE case. First, the reconstruction loss minimizes the reconstruction error of the overall architecture in the image space.
It is computed as the mean squared error between the predicted output $x^\prime$ and the ground truth $y$:
\begin{equation}
\label{eq:nca_training_reconstruction_loss}
    \mathcal{L}_{\text{REC}_{\text{NCA}}}(y, x^\prime) = \text{MSE}(y, x^\prime)\text{.}
\end{equation}
The latent loss plays a similar role to $\mathcal{L}_{\text{REC}_{\text{NCA}}}$ but in the latent space.
It enforces the similarity between the latent tensor of the ground truth, $\hat{y}$, and the one reconstructed by the NCA processing, $\hat{x}^\prime$, starting from the associated corrupted input:
\begin{equation}
\label{eq:nca_training_latent_loss}
    \mathcal{L}_{\text{LAT}}(\hat{y}, \hat{x}^\prime) = \text{MSE}(\hat{y}, \hat{x}^\prime)\text{.}
\end{equation}
Finally, the overflow loss regularizes the output magnitude of the NCA using an L1 norm.
As in~\cite{tesfaldet2022}, it is applied to both the output and the hidden channels of the NCA output tensor $\hat{x}^\prime$:
\begin{equation}
\label{eq:nca_training_overflow_loss}
    \mathcal{L}_{\text{OVER}}(\hat{x}^\prime) = \frac{1}{C} \big\|\hat{x}^\prime - \text{clip}(\hat{x}^\prime)\big\|_1\text{,}
\end{equation}
where $C$ is the number of channels considered.
Output channels are clipped in the interval $[0,1]$, while hidden channels are in the interval $[-1,1]$.

\section{Experiments and Results}
\label{se:experiments_and_results}

In this section, we evaluate the LNCA architecture in the IR task against several SotA models. In addition to the restoration performance, the analyses focus on time and memory resource efficiency, to assess the applicability of LNCA compared with existing NCA solutions.

\subsection{Datasets}
\label{subsubse:datasets}

This section collects the information related to the datasets involved in our experiments. To allow for a comprehensive performance evaluation, we adopted both synthetic and real-world datasets. Concerning synthetic datasets, we started from CelebA~\cite{liu2015}, CIFAR-10~\cite{krizhevsky2009}, and TinyImageNet~\cite{olga2015} and applied specific corruption patterns.
For denoising, we applied an additive Gaussian noise in a channel-coherent fashion.
This procedure generates tone shifts rather than completely out-of-context pixel values, making the task harder.
For deblurring, we use a camera motion blur.
It generates a defocus along a random direction coherent for all the elements in the image. 

For the real datasets, instead, we selected some of the most widely acknowledged in the image restoration field, namely Renoir~\cite{anaya2018} and SID~\cite{chen2018} for denoising, and GoPro~\cite{nah2017} and RealBlur~\cite{rim2020} for deblurring. 

Synthetic datasets grant fine-grained control over the type of corruption applied to the images, while real-world data allows for performance assessment in realistic scenarios. Also, synthetic and real datasets differ in their cardinality. On the one side, synthetic datasets contain many low-resolution images, while real datasets are composed of a few high-resolution images. Therefore, we need to apply different preprocessing to use these datasets properly. Specifically, for all the real datasets, we applied a tile-shrink-clean procedure, which converts the original image into a $n\times m$ patchwork, resizes each tile to $32\times32$ pixels, and then performs a cleaning step to remove redundant images. Conversely, for the synthetic datasets, we directly resized each image to the desired shape. 

\subsection{Training Procedure}
\label{subsubse:training_procedure}

In the following, we describe the experimental procedure adopted to collect the restoration performance and the time and memory requirements of LNCA and the baseline models from the literature. 
Specifically, we decided to use Restormer~\cite{zamir2022} and NAFNet~\cite{chen2022} as representatives of classic deep learning architectures, and ViTCA~\cite{tesfaldet2022} as the NCA flagship model.
On our side, as stated in Section~\ref{subsubse:neural_cellular_automata}, we propose two versions of the LNCA architecture, with a different NCA structure. The first version, referred to as LatentViTCA, adopts ViTCA as the NCA module. Conversely, the second version, called LatentNAFCA, has an attention-free NCA core, leading to a more lightweight architecture.

To keep the tests as fair as possible, we kept consistent training procedures across all the models. Specifically, we applied an 80-20\% split to obtain training and test sets followed by another 80-20\% split on the training set to extract the validation set.
When training on synthetic datasets, we employ curriculum learning in the training set, feeding images of increasing corruption, while validation and test sets use a fixed difficulty.
For real datasets, instead, we do not apply any curriculum learning. 

Each model has been trained for $20$ epochs, reaching a performance plateau. We opted for using Adam as an optimizer with cosine annealing scheduling~\cite{loshchilov2017} for the learning rate.
Furthermore, regarding CA-based architectures, we used a pool size of $1024$ and a hidden-channel size of $32$, as in~\cite{tesfaldet2022}.
During training, the number of CA iterations was uniformly sampled between $8$ and $32$, while for validation and testing, we fixed it to $64$.

\subsection{Results on Restoration Performance}
\label{subse:task_results}

This section comments on the results obtained by LNCA and the SotA IR counterparts concerning pure restoration performance. The goal of our proposed latent models is to improve the computational and memory efficiency of NCA architectures at the expense of task utility. Therefore, we expect latent architectures to perform slightly worse than the alternatives in pure restoration performance.

To assess performance, we adopt the reference metric in IR, namely the Structural Similarity Index Measure~\cite{zhou2004} (SSIM).
SSIM is a perception-based metric incorporating the concepts of luminance ($l$), contrast ($c$), and structure ($s$), which are estimated as follows:
\begin{equation}
\label{eq:metric_ssim_components}
\begin{split}
    l(x,y)=\frac{2\mu_x\mu_y+c_1}{\mu_x^2+\mu_y^2+c_1}\\
    c(x,y)=\frac{2\sigma_x\sigma_y+c_2}{\sigma_x^2+\sigma_y^2+c_2}\\
    s(x,y)=\frac{2\sigma_{xy}+c_3}{\sigma_x\sigma_y+c_3}\text{,}
\end{split}
\end{equation}
where $\mu_x$ and $\mu_y$ are the sample means, $\sigma_x$ and $\sigma_y$ the standard deviations, $\sigma_{xy}$ the cross-correlation, $c_1$, $c_2$, and $c_3$ stabilization coefficients.
These terms are then weighted (using $\alpha$, $\beta$ and $\gamma$ exponents) and combined to obtain the final formulation:
\begin{equation}
\label{eq:metric_ssim}
    SSIM(x,y)=l(x,y)^\alpha c(x,y)^\beta s(x,y)^\gamma\text{.}
\end{equation}
This metric leverages a kernel to extract the structural information of images, rather than focusing on single-pixel errors.
This leads to a more robust metric compared to other alternatives like MSE and PSNR.
 
\subsubsection{Denoising}
\label{subsubse:denoising}

\begin{table}[t!]
    \centering
    \caption{Denoising results on both synthetic (CelebA, CIFAR-10, and TinyImageNet) and real (Renoir and SID) datasets, reporting the SSIM obtained by each of the analyzed models. The best results are highlighted in bold, while the second best are underlined.}
    \label{table:denoising}
    \resizebox{.99\columnwidth}{!}{%
    \begin{tabular}{@{}lccccc@{}}
        \toprule
        Models & CelebA & CIFAR-10 & TinyImageNet & Renoir & SID \\
        \midrule
        NAFNet & \underline{0.958} & \textbf{0.924} & 0.837 & \underline{0.962} & \underline{0.752} \\
        Restormer & \textbf{0.967} & \underline{0.918} & \textbf{0.880}& \textbf{0.965} & \textbf{0.781} \\
        ViTCA & 0.942 & 0.910 & \underline{0.852} & 0.939 & 0.647 \\
        \midrule
        LatentViTCA & 0.889 & 0.855 & 0.802 & 0.877 & 0.646 \\
        LatentNAFCA & 0.837 & 0.833 & 0.776 & 0.880 & 0.652 \\
        \bottomrule
    \end{tabular}
    }
\end{table}

The first set of experiments focuses on the image-denoising task. Examining the results presented in Table~\ref{table:denoising}, NAFNet and Restormer achieve the highest denoising performance on all the tested datasets. Nevertheless, ViTCA obtains comparable results. As expected, latent models, namely LatentViTCA and LatentNAFCA, exhibit a decrease in performance due to the downsizing of the input. Upon close examination, we notice a decrease of $\approx5\%$ in denoising efficiency for LatentViTCA, with respect to ViTCA, and a slightly larger decrease of $\approx7\%$ for LatentNAFCA. We emphasize that a task utility downgrade is expected for latent architectures, as they are designed to trade restoration performance for computational and memory efficiency. In Section~\ref{subse:efficiency_results} we comment on this tradeoff by analyzing the resources required by latent models.

\subsubsection{Deblurring}
\label{subsubse:deblurring}

\begin{table}[t!]
    \centering
    \caption{Deblurring results on both synthetic (CelebA, CIFAR-10, and TinyImageNet) and real (GoPro and RealBlur) datasets, reporting the SSIM obtained by each of the analyzed models. The best results are highlighted in bold, while the second best are underlined.}
    \label{table:deblurring}
    \resizebox{.99\columnwidth}{!}{%
    \begin{tabular}{@{}lccccc@{}}
        \toprule
        Models & CelebA & CIFAR-10 & TinyImageNet & GoPro & RealBlur \\
        \midrule
        NAFNet & \underline{0.986} & \underline{0.970} & \underline{0.962} & \underline{0.908} & \underline{0.869} \\
        Restormer & \textbf{0.993} & \textbf{0.977} & \textbf{0.975} & \textbf{0.934} & \textbf{0.918} \\
        ViTCA & 0.941 & 0.903 & 0.870 & 0.773 & 0.814 \\
        \midrule
        LatentViTCA & 0.769 & 0.752 & 0.672 & 0.739 & 0.772 \\
        LatentNAFCA & 0.745 & 0.739 & 0.673 & 0.739 & 0.771 \\
        \bottomrule
    \end{tabular}
    }
\end{table}

Results on the deblurring task, follow a similar trend to denoising, as shown in Table~\ref{table:deblurring}. In particular, NAFNet and Restormer still achieve the best deblurring results. ViTCA follows as the third best alternative, although the performance gap is higher than the denoising task. Compared to ViTCA, latent models exhibit a greater decrease in performance of $\approx13\%$ for LatentViTCA and $\approx15\%$ for LatentNAFCA.
This gap is less prominent in real-world datasets~-- GoPro and RealBlur~-- which are affected by less critical corruptions than synthetic datasets.

\subsection{Efficiency Results}
\label{subse:efficiency_results}

This section focuses on the efficiency analysis of the proposed solution, LNCA, compared with the existing IR architectures. We measure and compare three metrics: training memory usage, training latency, and inference latency. These metrics summarize the resource usage and computational requirements of a generic deep learning architecture.
The objective of LNCA is to reduce these metrics with respect to other NCA-based models. 
In this context, our primary comparison is with ViTCA, which belongs to the NCA family. 
However, we will also extend our tests to NAFNet and Restormer to ensure a more comprehensive understanding of the results.

To enable testing on a wider range of input sizes, the following experiments were performed on a cloud service using an Nvidia A100 GPU with $80$GB of VRAM, which has been the main limiting factor for NCA architectures.

\subsubsection{Training memory requirement}
\label{subsubse:training_memory_requirement}

\begin{table}[t!]
    \centering
    \caption{Training memory requirements tests.
    Results are reported once the memory requirements have stabilized after an initial warm-up phase, measured in GB.
    If the configuration requires more than the available $\mathbf{80}$ GB of VRAM, the measurement is not available and is indicated by "--".
    The best results are highlighted in bold, while the second best are underlined.}
    \label{table:training_memory}
    \resizebox{.99\columnwidth}{!}{
    \begin{tabular}{@{}lccccccccc@{}}
        \toprule
        \multirow{2}{*}{Models} & \multicolumn{3}{c}{$32\times32$} & \multicolumn{3}{c}{$64\times64$} & \multicolumn{3}{c}{$128\times128$} \\
        \cmidrule{2-10}
        & $8$ & $16$ & $32$ & $8$ & $16$ & $32$ & $8$ & $16$ & $32$ \\
        \midrule
        NAFNet & 4.1& 4.1 & 4.2 & 4.2 & 8.2 & 8.8 & 8.7 & 19.6 & 35.0 \\
        Restormer & 4.4 & 4.4 & 8.5 & 8.8 & 16.8 & 32.9 & 32.8 & 65.6 & -- \\
        ViTCA & 8.7 & 16.9 & 33.8 & 33.8 & 65.6 & -- & -- & -- & -- \\
        \midrule
        LatentViTCA & \underline{0.5} & \underline{1.1} & \underline{1.2} & \underline{2.2} & \underline{4.4} & \underline{4.7} & \underline{8.7} & \underline{16.9} & \underline{17.4} \\
        LatentNAFCA & \textbf{0.5} & \textbf{0.5} & \textbf{1.1} & \textbf{1.1} & \textbf{2.1} & \textbf{4.1} & \textbf{4.1} & \textbf{8.2} & \textbf{16.4} \\
        \bottomrule
    \end{tabular}}
\end{table}
\begin{figure}
    \centering
    \includegraphics[width=\columnwidth]{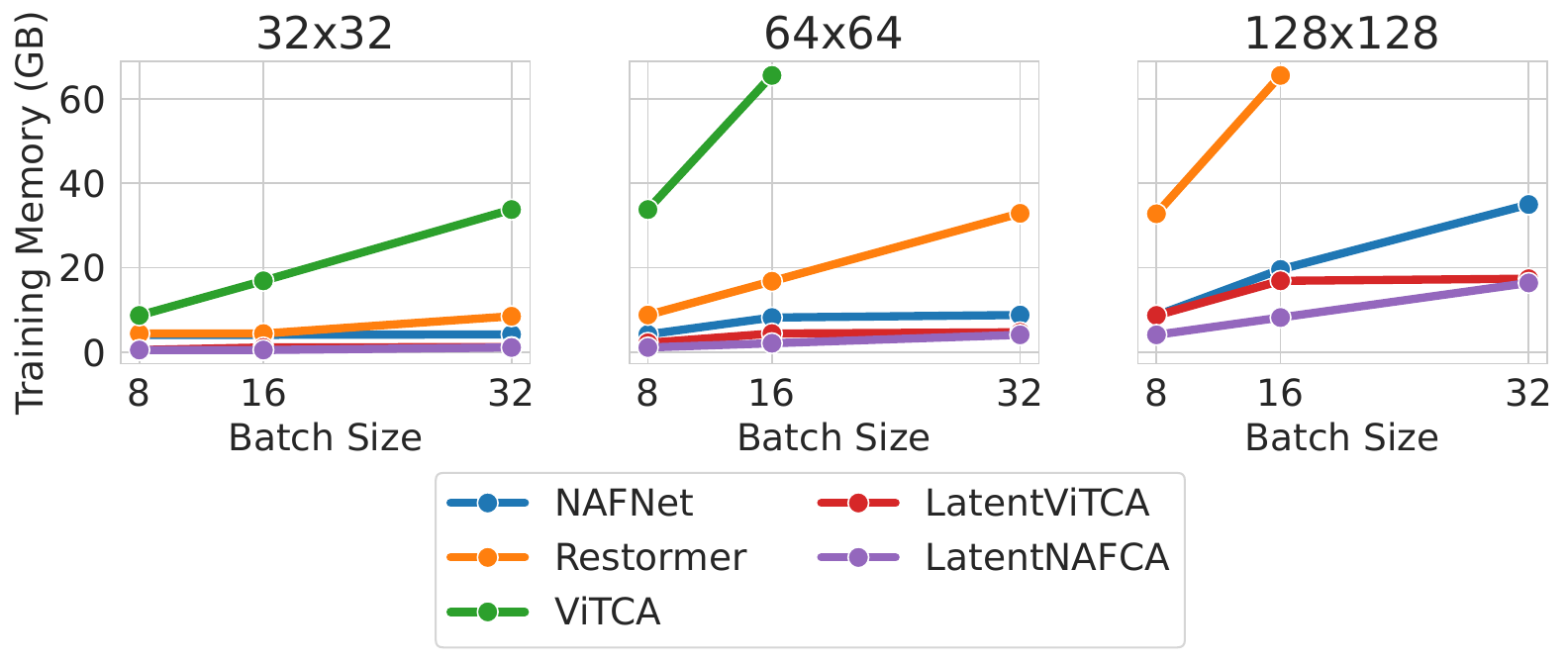}
    \caption[Training memory trend]{Training memory trend of the tested models across the different configurations.}
    \label{fig:training_memory_trend}
\end{figure}

This metric (Table~\ref{table:training_memory} and Figure~\ref{fig:training_memory_trend}) is calculated as the highest VRAM requirement for each model at a given input resolution, typically reached after an initial warm-up period and then held constant throughout training.

Starting from the lowest image resolution tested, namely $32\times32\times3$ with a batch size of 8, we notice a reduction of $\approx94\%$ for both our solutions compared to ViTCA.
This reduction remains nearly constant across all input configurations.
Furthermore, considering the minimum memory requirement of ViTCA, i.e., $8.7$GB, we achieve a $16$x improvement in the training input size by using its latent counterpart.
Interestingly, the latent models are also up to $\approx88\%$ more efficient than the classical solutions, which should have a natural advantage due to their non-recurrent structure.
Overall, we can see that our solutions (and NAFNet) cope well with the increase in batch size, keeping the requirements almost stable, while the other models turn out to be less scalable.

\subsubsection{Training latency}
\label{subsubse:training_latency}

In this experiment, we adopt a dummy dataset of $1000$ noisy samples ($80\%$ training, $20\%$ validation) and perform a $10$-epoch training.
We compute the average training latency discarding the values of the initial warm-up epoch.
For latent models, we sum the AE and NCA training times to get a fair estimate of total training time.

\begin{figure}[t]
    \centering
    \includegraphics[width=\columnwidth]{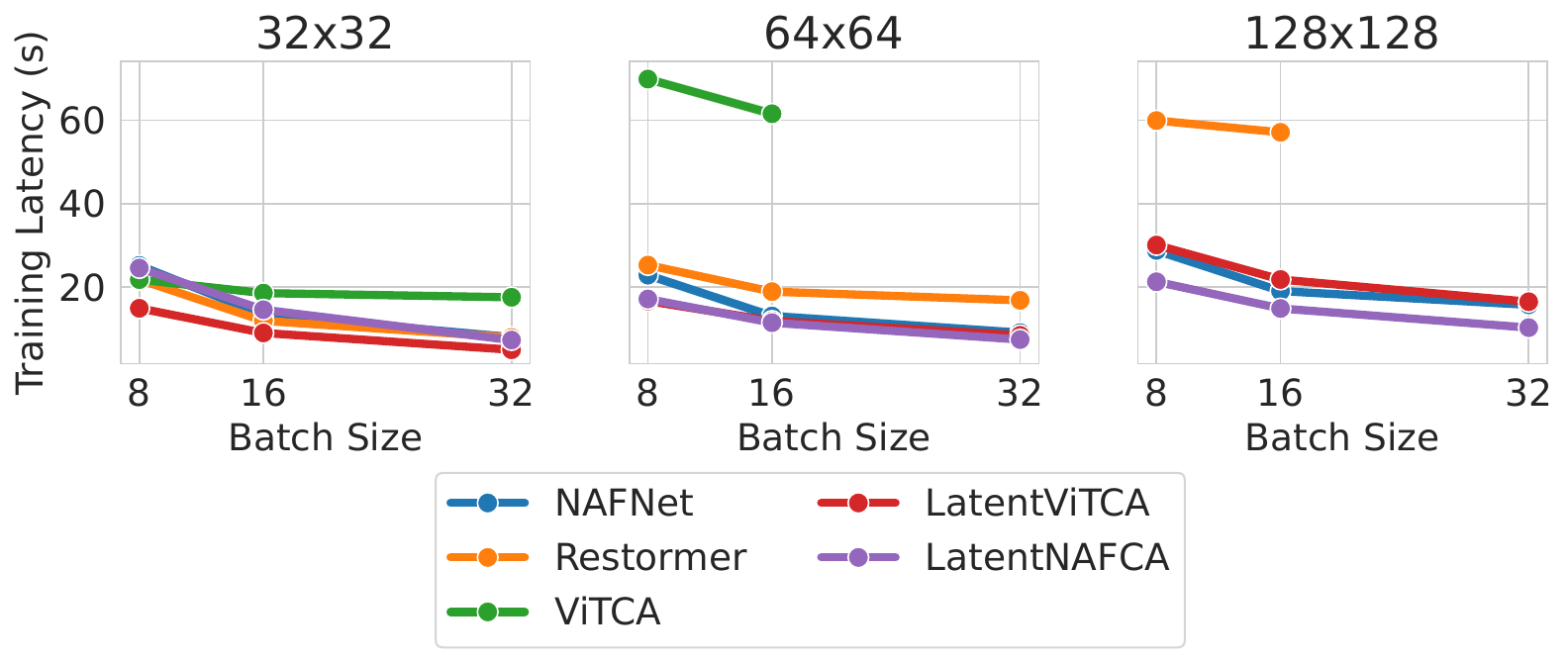}
    \caption[Training latency trend]{Training latency trend of the tested models across the different configurations.}
    \label{fig:training_latency_trend}
\end{figure}

As expected, with a fixed-size dataset, increasing the batch size leads to an improvement in the overall training time for all models (Table~\ref{table:training_latency} and Figure~\ref{fig:training_latency_trend}).
At the lowest resolution ($32\times32\times3$), the difference between ViTCA and the latent models is negligible, although the decreasing slope is less pronounced in the former. 
Increasing the resolution, we notice the degradation in performance of ViTCA and, subsequently, of Restormer, while the remaining models keep stable timings.
In our view, the point at which performance begins to decline is closely tied to the memory usage of the models. 
Specifically, processing latency deteriorates as the model starts to saturate the GPU VRAM.
Overall, latent models achieve up to $\approx80\%$ reduction in training latency when compared to ViTCA, making them comparable to the NAFNet.

\begin{table}[t!]
    \centering
    \caption{Training latency tests performed with on a dummy dataset of $\mathbf{1000}$ samples ($\mathbf{80-20}$ split).
    The results are the average of $\mathbf{10}$ training epochs (after the initial warm-up), measured in seconds.
    If the configuration requires more than the available $\mathbf{80}$ GB of VRAM, the measurement is not available and is indicated by "--".
    The best results are highlighted in bold, while the second best are underlined.}
    \label{table:training_latency}
    \resizebox{.99\columnwidth}{!}{
    \begin{tabular}{@{}lccccccccc@{}}
        \toprule
        \multirow{2}{*}{Models} & \multicolumn{3}{c}{$32\times32$} & \multicolumn{3}{c}{$64\times64$} & \multicolumn{3}{c}{$128\times128$} \\
        \cmidrule{2-10}
        & $8$ & $16$ & $32$ & $8$ & $16$ & $32$ & $8$ & $16$ & $32$ \\
        \midrule
        NAFNet & 25.1 & 13.3 & 7.9 & 22.8 & 13.0 & 9.0 & \underline{28.8} & \underline{19.0} & \underline{15.7} \\
        Restormer & \underline{21.7} & \underline{11.9} & 7.9 & 25.2 & 18.9 & 16.8 & 59.9 & 57.1 & -- \\
        ViTCA & 21.7 & 18.5 & 17.5 & 69.9 & 61.6 & -- & -- & -- & -- \\
        \midrule
        LatentViTCA & \textbf{14.9} & \textbf{8.9} & \textbf{4.9} & \textbf{16.6} & \underline{11.8} & \underline{8.4} & 30.0 & 21.8 & 16.4 \\
        LatentNAFCA & 24.5 & 14.5 & \underline{7.2} & \underline{17.1} & \textbf{11.4} & \textbf{7.4} & \textbf{21.2} & \textbf{14.8} & \textbf{10.2} \\
        \bottomrule
    \end{tabular}}
\end{table}

\subsubsection{Inference latency}
\label{subsubse:inference_latency}

Regarding the inference latency, tests are performed using a single batch as input and averaging over $10$ iterations.
Similarly to the training latency, we discard the first processing of each model as a warm-up.

\begin{figure}
    \centering
    \includegraphics[width=\columnwidth]{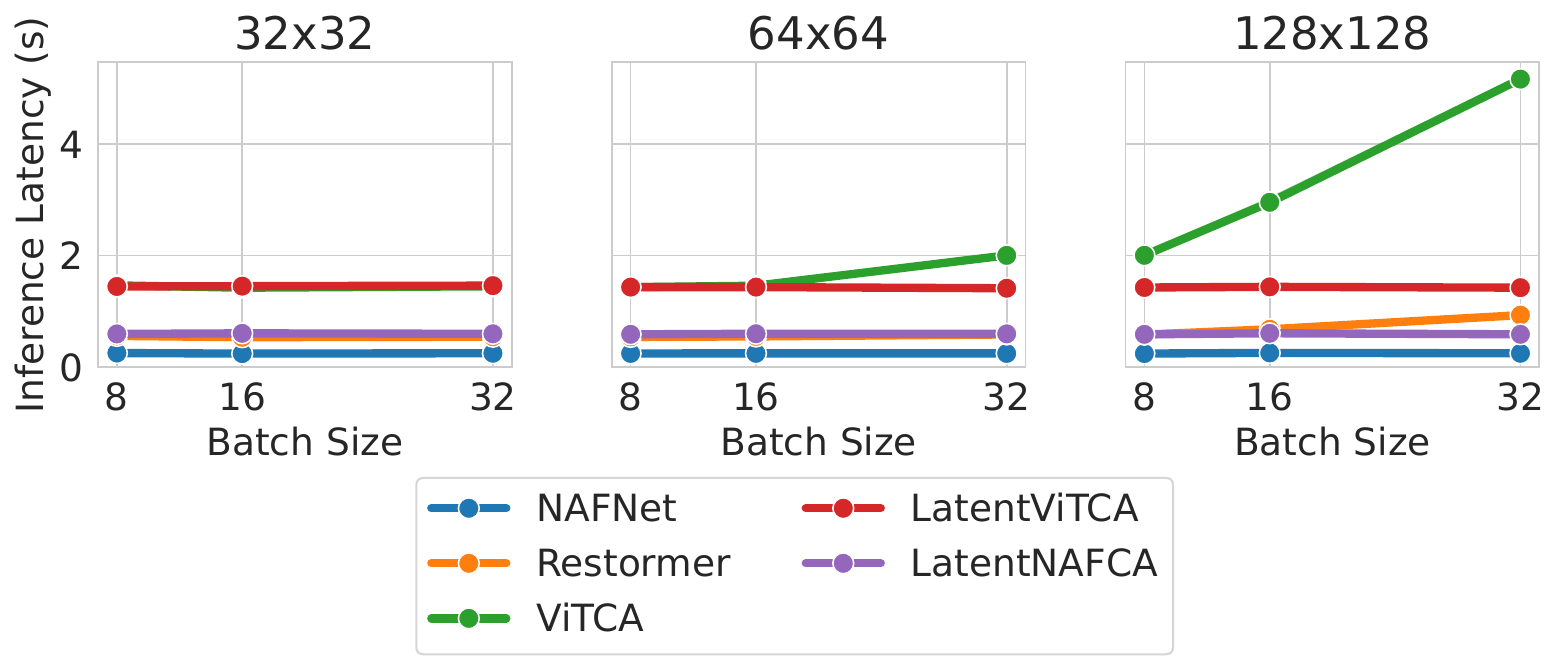}
    \caption[Inference latency trend]{Inference latency trend of the tested models across the different configurations.}
    \label{fig:inference_latency_trend}
\end{figure}

We expect our solutions and ViTCA to be significantly slower than classical models due to the NCA computation.
In fact, we fix these models to perform $t=64$ NCA processing steps instead of the single pass used by the non-NCA models.
Looking at the results (Table~\ref{table:inference_latency} and Figure~\ref{fig:inference_latency_trend}), we can see that our solutions and NAFNet maintain almost stable performance across all resolutions, while the other models diverge once a memory breakpoint is reached.
As mentioned in the previous section, we believe that this trend is directly related to VRAM usage.
At the highest resolution ($128\times128\times3$), where memory requirements are demanding, our latent models are respectively up to $\approx72\%$ and $\approx89\%$ faster than ViTCA, with LatentNAFCA also $\approx37\%$ faster than Restormer.

\begin{table}[t!]
    \centering
    \caption{Inference latency tests performed on a single batch.
    The results are the average of $\mathbf{10}$ inference steps (discarding the initial warm-up), measured in seconds.
    The best results are highlighted in bold, while the second best are underlined.}
    \label{table:inference_latency}
    \resizebox{.99\columnwidth}{!}{
    \begin{tabular}{@{}lccccccccc@{}}
        \toprule
        \multirow{2}{*}{Models} & \multicolumn{3}{c}{$32\times32$} & \multicolumn{3}{c}{$64\times64$} & \multicolumn{3}{c}{$128\times128$} \\
        \cmidrule{2-10}
        & $8$ & $16$ & $32$ & $8$ & $16$ & $32$ & $8$ & $16$ & $32$ \\
        \midrule
        NAFNet & \textbf{0.25} & \textbf{0.24} & \textbf{0.25} & \textbf{0.24} & \textbf{0.24} & \textbf{0.24} & \textbf{0.24} & \textbf{0.25} & \textbf{0.25} \\
        Restormer & \underline{0.56} & \underline{0.53} & \underline{0.54} & \underline{0.54} & \underline{0.55} & \underline{0.58} & \underline{0.58} & 0.68 & 0.93 \\
        ViTCA & 1.47 & 1.42 & 1.45 & 1.43 & 1.46 & 2.00 & 2.01 & 2.96 & 5.17 \\
        \midrule
        LatentViTCA & 1.45 & 1.45 & 1.46 & 1.43 & 1.43 & 1.41 & 1.43 & 1.44 & 1.42 \\
        LatentNAFCA & 0.59 & 0.60 & 0.60 & 0.59 & 0.60 & 0.59 & 0.59 & \underline{0.60} & \underline{0.59} \\
        \bottomrule
    \end{tabular}}
\end{table}

\section{Conclusions}
\label{se:conclusions}

NCAs are fascinating models capable of learning complex system behavior from simple local interactions. Yet, their practical application has been severely limited by their computational constraints. This work focused on mitigating this critical challenge, providing an NCA model, called LNCA, with low memory requirements and latency bottlenecks. LNCA is a novel deep learning-based architecture that moves the computationally demanding NCA computation from the input space to a lower-dimensional manifold. This manifold is tailored by an autoencoder to extract the essential information for the NCA to solve the task. We validated LNCA in the field of image restoration, specifically addressing denoising and deblurring tasks. The experiments showcased an effective reduction in the computational requirements of NCA. With respect to the SotA, LNCA achieved up to a $16\times$ improvement in the maximum input size, and up to a $\approx94\%$ and $\approx80\%$ reductions in memory requirements and processing latency, respectively, with the drawback of an overall degradation in restoration performance of $\approx10\%$. 
In conclusion, LNCA successfully addressed some of the inherent limitations of NCA models, providing a flexible and practical solution for the integration of cellular automata into deep learning architectures.


\bibliographystyle{unsrt}  
\bibliography{references}
\end{document}